\documentclass[12pt,aps,prd]{revtex4}
\begin{document}

%\title{Polyakov Loops, Z(N) Symmetry, and Sine-Law Scaling}
\title{Polyakov Loop Models, Z(N) Symmetry, and Sine-Law Scaling}
 
\author{Peter N. Meisinger}
\author{Michael C. Ogilvie}

\affiliation{Dept. of Physics, Washington University, St. Louis, MO 63130 USA}

\begin{abstract}
We construct an effective action for Polyakov loops using the
eigenvalues of the Polyakov loops as the fundamental variables. We assume $%
Z(N)$ symmetry in the confined phase, a finite difference in energy
densities between the confined and deconfined phases as $T\rightarrow 0$,
and a smooth connection to perturbation theory for large $T$. The
low-temperature phase consists of $N-1$ independent fields fluctuating around an
explicitly $Z(N)$ symmetric background. In the low-temperature phase, the
effective action yields non-zero string tensions for all representations
with non-trivial $N$-ality. Mixing occurs naturally between representations
of the same $N$-ality. Sine-law scaling emerges as a special case,
associated with nearest-neighbor interactions between Polyakov loop
eigenvalues.

\end{abstract}

\maketitle

%Effective actions for the Polyakov loop are directly relevant to the phase
%diagram and equation of state for QCD and related theories. Furthermore,
%there may be a close relationship between the correct effective action and
%the underlying mechanisms of confinement. In this letter, we construct a
%general effective action for the Polyakov loop, making as few assumptions as
%possible. Our goal is a comprehensive model which describes both the
%confined and deconfined phases of $SU(N)$ pure gauge theories at finite
%temperature, as well as the properties of the deconfining phase transition.
%We will describe below the general form of such an effective action, using
%the Polyakov loop eigenvalues as natural variables.
%Effective actions of
%this type have previously been constructed for
%QCD at high temperature \cite{Gross:1980br,Weiss:1980rj,Weiss:1981ev}, 
%QCD in two dimensions\cite{Minahan:1993np,Polychronakos:1999sx},
%and for lattice gauge theories at
%strong coupling\cite{Polonyi:1982wz,Ogilvie:1983ss,Green:1983sd,Drouffe:1984hb}.

Effective actions for the Polyakov loop are directly relevant to the phase
diagram and equation of state for QCD and related theories
\cite{Pisarski:2000eq,Dumitru:2001xa,Meisinger:2001cq,Dumitru:2003hp,Meisinger:2003id}. 
There is also good reason to believe that Polyakov loop effects play an important 
role in chiral symmetry restoration
\cite{Gocksch:1984yk,Dumitru:2000in,Fukushima:2002ew,Fukushima:2003fm,Mocsy:2003qw}.
Furthermore,
there may be a close relationship between the correct effective action and
the underlying mechanisms of confinement
\cite{Meisinger:1997jt,Meisinger:2002ji,Kogan:2002yr,Mocsy:2003tr,Mocsy:2003un}.

In this letter, we construct a
general effective action for the Polyakov loop, making as few assumptions as
possible. Our goal is a comprehensive model which describes both the
confined and deconfined phases of $SU(N)$ pure gauge theories at finite
temperature, as well as the properties of the deconfining phase transition.
We will describe below the general form of such an effective action, using
the Polyakov loop eigenvalues as natural variables.
Effective actions of
this type have previously been constructed for
QCD at high temperature \cite{Gross:1980br,Weiss:1980rj,Weiss:1981ev}, 
QCD in two dimensions\cite{Minahan:1993np,Polychronakos:1999sx},
and for lattice gauge theories at
strong coupling\cite{Polonyi:1982wz,Ogilvie:1983ss,Green:1983sd,Drouffe:1984hb}.

The Polyakov loop $P(x)$ is the natural
order parameter for the deconfinement transition. It is defined as a
path-ordered exponential $P(x)=\emph{P}\exp \left[ i\int_{0}^{\beta }d\tau
A_{0}\left( \tau ,x\right) \right] $ where $x\,$is a $d$-dimensional spatial
vector and $\beta \,$is the inverse temperature. The confined phase has a $%
Z(N)$ symmety that implies $\left\langle Tr_{F}\,\,P^{k}\left( x\right)
\right\rangle =0$ for all $n$ not divisible by $N$: $k|N\neq 0$. For these
powers of the Wilson line, the asymptotic behavior 
\begin{equation}
\left\langle Tr_{F}P^{k}\left( x\right) Tr_{F}P^{+k}\left( y\right)
\right\rangle \propto \exp \left[ -\beta \sigma _{k}\left| x-y\right|
\right] 
\end{equation}
is observed in lattice simulatiosn as $\left| x-y\right| \rightarrow \infty $%
. The string tension $\sigma _{k}$ is in general temperature-dependent.
Similar behavior is also seen for the two point functions associated with
irreducible representations of the gauge group $\left\langle Tr_{R}P\left(
x\right) Tr_{R}P^{+}\left( y\right) \right\rangle $. Every irreducible
representation $R$ has an associated $N$-ality $k_{R}$ such that $%
Tr_{R}P\rightarrow z^{k_{R}}Tr_{R}P$ under a \ \ global $Z(N)$
transformation, with $k_{R}\in \{0,..,N-1\}$ and $z\in Z(N)$. Any
representation with $k_{R}\neq 0$ gives an operator $Tr_{R}P$ which is an
order parameter for the spontaneous breaking of the global $Z(N)$ symmetry.

It is widely held that the asymptotic string tension depends only on the $N$%
-ality of the representation. In fact, lattice data from simulations of four
dimension $SU(3)$ gauge theory do not yet show this behavior \cite{Deldar:1999vi,Bali:2000un}. 
Instead,
the string tension $\sigma _{R}$ associated with a representation $R$ scales
approximately as 
\begin{equation}
\sigma _{R}=\frac{C_{R}}{C_{F}}\sigma _{F} 
\end{equation}
where $C_{R}$ is the quadratic Casimir invariant for the representation $R$. 
This behavior can be rigourously
demonstrated in two-dimensional gauge theories. For simplicity, we will
refer to this as Casimir scaling. We use the term $Z(N)$ scaling to descibe
the asymptotic behavior

\begin{equation}
\sigma _{R}=\frac{k\left( N-k\right) }{N-1}\sigma _{F} 
\end{equation}
where $k$ is the $N$-ality of the representation $R$. As we will show below, 
$Z(N)$ scaling is obtained from Casimir scaling at large distances if there
is mixing between representations of the same $N$-ality. If $Z(N)$ scaling
holds, there are $\lbrack N/2\rbrack $ distinct asymptotic string tensions,
where $\lbrack N/2\rbrack $ is \ the largest integer less than or equal to $%
N/2$. For the first $\left[ N/2\right] $ antisymmetric representations made
by stacking boxes in Young tableux, Casimir and $Z(N)$ scaling are
identical. Another possible scaling law for the string tensions is sine-law
scaling

\begin{equation}
\sigma _{R}=\frac{\sin \left( \frac{\pi k}{N}\right) }{\sin \left( \frac{\pi 
}{N}\right) }\sigma _{F} 
\end{equation}
which has been shown to occur in softly broken $\emph{N}=2$ super Yang-Mills
theories\cite{Douglas:1995nw}
and in MQCD\cite{Hanany:1997hr}.

In a gauge in which $A_{0}$ is time independent and diagonal, we may write $%
P $ in the fundamental representation as 
\begin{equation}
P_{jk}=\exp \left( i\theta _{j}\right) \delta _{jk} 
\end{equation}
where we shall refer to the $N\,$\ numbers $\theta _{j}$ as the eigenvalues.
They are not independent because $\det \left( P\right) =1$ implies 
\begin{equation}
\sum_{j}\theta _{j}=0\,mod\,2\pi \rm{.} 
\end{equation}
The information in the different representations is redundant. All the
information is contained in the $N-1$ independent eigenvalues of $P$.

Without spatial gauge fields, the only gauge-invariant operators we can
construct are class functions of $P$, depending solely on its eigenvalues.
Thus the effective action should take $P$ to be in the Cartan, or maximally
commuting, subgroup $U(1)^{N-1}$. An effective action constructed from $%
SU(N) $ matrices would introduce spurious Goldstone bosons associated with
the off-diagonal components of $P$ in the deconfined phase. On the other
hand, the simplest Landau-Ginsburg models of the $SU(2)$ and $SU(3)\,$%
deconfining transitions have used $Tr_{F}P$ as the basic field. These two
cases are special because $Tr_{F}P$ specifies the Polyakov loops in all
other representations. In $SU(4)$, there are sets of eigenvalues for which 
\begin{equation}
Tr_{F}P=0 \quad\rm{and}\quad Tr_{F}P^{2}=0 
\end{equation}
consistent with a $Z(4)$ symmetry, and another set for which 
\begin{equation}
Tr_{F}P=0 \quad\rm{and}\quad Tr_{F}P^{2}\neq 0 
\end{equation}
consistent with a $Z(2)$ symmetry. Thus knowledge of $Tr_{F}P$ alone is
insufficient for $N>3$\cite{Meisinger:2001cq}. From
the characteristic polynomial, one may show that the eigenvalues of a
special unitary matrix are determined by the set $\left\{
Tr_{F}P^{k}\right\} $ with $k=1..N-1$, and of course \textit{vice versa}.

\vspace{1pt}In both the confined and deconfined phases, we would like our
effective theory to proceed from a classical field configuration which has
the symmetries of the phase. If we denote that field cofiguration as $P_{0}$
in the confined phase, $Z(N)\,$symmetry requires that $Tr_{F}P_{0}^{k}=0$
for all $k$ not divisible by $N$. Enforcing this requirement for $k=1$ to $%
N-1$ leads to a unique set of eigenvalues via the characteristic equation $%
z^{N}+\left( -1\right) ^{N}=0$, but it is instructive to derive the set
another way. \vspace{1pt}For temperatures $T$ below the deconfinement
transition $T_{d}$, center symmetry is unbroken. Unbroken center symmetry
implies that $zP_{0}$ is equivalent to $P_{0}$ after an $SU(N)$
transformation:

\begin{equation}
zP_{0}=gP_{0}g^{+}\rm{.} 
\end{equation}
This condition in turn implies $Tr_{R}P_{0}=0$ for all representations $R$
with non-zero $N$-ality, which means that all representations with non-zero $%
N$-ality are confined. The most general form for $P_{0}$ may be given as $%
hdh^{+}$, where $h\in SU(N)$, and $d$ is the diagonal element of $SU(N)$ of
the form 
\begin{equation}
d=w\,\,diag\left[ z,z^{2},..,z^{N}=1\right] 
\end{equation}
where $z$ is henceforth $\exp \left( 2\pi i/N\right) $, the generator of $%
Z(N)$. The phase $w$ ensures that $d$ has determinant $1$, and is given by $%
w=\exp \left[ -\left( N+1\right) \pi i/N\right] $. Strictly speaking, $w$ is
required only for $N$ even, but it is convenient to use it consistently. We
will henceforth identify $P_{0}$ with $d$. Another useful representation is $%
\left( P_{0}\right) _{jk}=\delta _{jk}\exp \left[ i\theta _{j}^{0}\right] $
where 
\begin{equation}
\theta _{j}^{0}==\frac{\pi }{N}\left( 2j-N-1\right) \rm{.} 
\end{equation}
Thus the $Z(N)$-symmetric arrangement of eigenvalues is uniform spacing
around the unit circle. This is consistent with the known large-$N$\
behavior of soluble models in the confined phase\cite{Aharony:2003sx}.

\vspace{1pt}Assume that $P_{0}$ is the global minimum of the potential $V$
associated with the effective action for temperatures less than the
deconfining temperature $T_{d}$. Because the gauge fields transform as the
adjoint representation, $V$ is a class function depending only on
representations of zero $N$-ality. It is thus a function only of the
differences in eigenvalues $\theta _{j}-\theta _{k}$, with complete
permutation symmetry as well. In the low-temperature, confining phase, we
consider small fluctuations about $P_{0}$, defining $\theta _{j}=\theta
_{j}^{0}+\delta \theta _{j}$. 
Although this approximation may not be {\it a priori} valid,
the assumption that fluctuations are small
can be justified in the large-$N$ limit. 
For small fluctuations 
\begin{equation}
Tr_{F}P^{k}=\sum_{n=1}^{N}w^{k}z^{kn}e^{ik\delta \theta _{n}}\simeq
\sum_{n=1}^{N}w^{k}z^{kn}ik\,\delta \theta _{n} 
\end{equation}
provided $k$ is not divisible by $N$. Note that $Tr_{F}P^{k}$ takes the form
of a discrete Fourier transform in eigenvalue space. We define the Fourier
transform of the fields $\phi _{n}$ as 
\begin{equation}
\phi _{k}=\sum_{n=1}^{N}z^{kn}\delta \theta _{n} 
\end{equation}
so $Tr_{F}P^{k}=ikw^{k}\phi _{k}$. We will show below that the fields $\phi
_{k}$ are the normal modes of the $Z(N)$-symmetric phase in a quadratic
approximation. The mode $\phi _{N}\equiv \phi _{0}\,$\ is identically zero
for $SU(N)$, and from the reality of $\theta $, we have $\phi _{N-n}=\left(
\phi _{n}\right) ^{\ast }$. In the case where $k$ is divisible by $N$, $%
Tr_{F}P^{k}$ has a leading constant behavior of $Nw^{k}$, and the term
linear in $\phi $ vanishes. The adjoint Polyakov loop operator is
approximately $Tr_{A}P=Tr_{F}P\,\,Tr_{F}P^{+}-1\simeq \phi _{1}\phi
_{1}^{\ast }-1$. The operator $\phi _{1}\phi _{1}^{\ast }$ has a non-zero
vacuum expectation value, and should couple to scalar glueball states.

Operators with the same $N$-ality generally give inequivalent expressions
when written in terms of the the $\phi $ variables. For example, $%
Tr_{F}P^{k}\propto \phi _{k}$ and $\left( Tr_{F}P\right) ^{k}\propto
\left( \phi _{1}\right) ^{k}$.\ Group characters $\chi _{R}(P)$ are
represented as sums of terms with the same $N$-ality, but with different
mode content. For example, in $SU(4)$, the $\mathbf{10}$ and $\mathbf{6}$
representations are given by 
\begin{equation}
\chi _{S,A}=\frac{1}{2}\left[ \left( Tr_{F}P\right) ^{2}\pm
Tr_{F}P^{2}\right] \simeq \frac{1}{2}\left[ \pm 2\phi _{2}+i\phi
_{1}^{2}\right] \rm{.} 
\end{equation}
As we discuss below in detail, these different combinations of fields will
in general produce several different excitations, and only the lightest
states will dominate at large distances.

The form of the effective action is fixed at high temperature by
perturbation theory. The form of the effective action at high temperatures
can be written as \cite{Bhattacharya:1990hk,Bhattacharya:1992qb}
\begin{equation}
S_{eff}=\beta \int d^{3}x\left[ T^{2}Tr_{F}\,\left( \nabla \theta \right)
^{2}+V_{1L}(\theta )\right].
\end{equation}
The kinetic term
is obtained from the underlying gauge action via $\frac{1}{2}Tr_{F}\,F_{\mu
\nu }^{2}\rightarrow \frac{1}{2}2Tr_{F}\,\left( \nabla A_{0}\right)
^{2}\rightarrow T^{2}Tr_{F}\,\left( \nabla \theta \right) ^{2}$. The
potential $V_{1L}(\theta )$ is obtained from one-loop perturbation theory.
For our purposes, it is conveniently expressed as \cite{Gross:1980br,Weiss:1980rj,Weiss:1981ev}
\begin{eqnarray*}
V_{1L}(\theta ) &=&-\sum_{n=1}^{\infty }\frac{2}{\pi ^{2}}\frac{T^{4}}{n^{4}}%
\left[ \left| TrP^{n}\right| ^{2}-1\right] \\
&=&-\sum_{n=1}^{\infty }\frac{2}{\pi ^{2}}\frac{T^{4}}{n^{4}}\left[
N-1+\sum_{j\neq k}\cos \left( n\left( \theta _{j}-\theta _{k}\right) \right)
\right].
\end{eqnarray*}
This series can
be summed to a closed form in terms of the 4th Bernoulli polynomial. The
complete one-loop expression has been obtained recently
\cite{Diakonov:2003yy,Diakonov:2003qb,Diakonov:2004kc};
the complete kinetic term has a $\theta $%
-dependent factor in front of the derivatives.

There are $N$ equivalent solutions of the form 
\begin{equation}
\theta _{j}^{(p)}=\frac{2\pi p}{N} 
\end{equation}
related by $Z(N)$ symmetry breaking. All of these solutions break $Z(N)$
symmetry, with $Tr_{F}P=N\exp \left( 2\pi ip/N\right) $.\thinspace\ For
these values of $\theta $, we recover the standard black-body result for the
free energy.

A sufficiently general form of the action at all temperatures has the form 
\begin{equation}
S_{eff}=\beta \int d^{3}x\left[ \kappa T^{2}Tr_{F}\,\left( \nabla \theta
\right) ^{2}+V(\theta )\right] 
\end{equation}
where $\kappa $ is a temperature dependent correction to the kinetic term,
and $V$ is a function only of the adjoint eigenvalues $\theta _{j}-\theta
_{k}$. More complicated derivative terms can be added as necessary.

We assume that there is a finite free energy density difference associated
with different values of $P$ as $T\rightarrow 0$. Because the eigenvalues
are dimensionless, this requires terms in the potential with coefficients
proportional to $\left( mass\right) ^{4}\,$as $T\rightarrow 0$. We can
expand the potential to quadratic order around $P_{0}$%
\begin{equation}
V\left( \theta \right) \simeq V(\theta ^{0})+\sum_{j,k}\frac{1}{2}\left[ 
\frac{\partial ^{2}V}{\partial \theta _{j}\partial \theta _{k}}\right]
_{\theta ^{0}}\delta \theta _{j}\delta \theta _{k} 
\end{equation}
where the coefficient in the expansion depends only on $\left| j-k\right| $. 
The quadratic piece is thus diagonalized by
Fourier transform, and we can write 
\begin{equation}
V\left( \theta \right) \simeq V(\theta ^{0})+\sum_{n=1}^{N-1}M_{n}^{4}\phi
_{n}\phi _{N-n}\rm{.} 
\end{equation}
Similarly, the kinetic term becomes 
\begin{equation}
\kappa T^{2}Tr_{F}\,\left( \nabla \theta \right) ^{2}=\frac{\kappa T^{2}}{N}%
\sum_{n=1}^{N-1}\left( \nabla \phi _{n}\right) \left( \nabla \phi
_{N-n}\right) \rm{.} 
\end{equation}
Once an ordering of eigenvalues is chosen, $Z(N)$ symmetry is expressed as a
discrete translation symmetry in eigenvalue space. If we write the
higher-order parts of $S_{eff}$ in terms of the Fourier modes $\phi _{n}$,
each interaction will respect global conservation of $N$-ality. For example,
in $SU(4)$, an interaction of the form $\phi _{1}^{2}\phi _{2}$ is allowed,
but not $\phi _{1}^{2}\phi _{2}^{2}$.

The confining behavior of Polyakov loop two-point functions at low
temperatures, for all representations of non-zero $N$-ality, is natural in
the effective model. If the interactions are neglected, we can calculate the
behavior of Polyakov loop two-point functions at low temperatures from the
quadratic part of $S_{eff}$. We have for large distances 
\begin{equation}
\left\langle Tr_{F}P^{n}\left( x\right) Tr_{F}P^{+n}\left( y\right)
\right\rangle \propto \left\langle \phi _{n}\left( x\right) \phi _{n}^{\ast
}\left( y\right) \right\rangle \propto \exp \left[ -\frac{\sigma _{n}}{T}%
\left| x-y\right| \right] 
\end{equation}
where $\sigma _{n}\left( T\right) =\sqrt{NM_{n}^{4}(T)/\kappa (T)}$ is
identified as the string tension for the $n$'th mode at temperature $T$.
Interactions may cause the physical string tensions to be significantly
different from the tree-level result, but the general field-theoretic
framework remains in any case. Of course, $\phi _{N-n}=\left( \phi
_{n}\right) ^{\ast }$ implies $\sigma _{n}\left( T\right) =\sigma _{N-n}(T)$%
. The number of different string tensions is $\left[ N/2\right] $, the
greatest integer less than or equal to $N/2$. The zero-temperature string
tension is given at tree level by 
\begin{equation}
\sigma _{n}^{2}\left( 0\right) =NM_{n}^{4}(0)/\kappa (0) 
\end{equation}

This formalism gives a natural mechanism for the transition from Casimir
scaling to scaling based on $N$-ality at large distances. As we have noted
above, there are many composite operators with the same $N$-ality. For
example, in $SU(8)$, the operators $\phi _{1}^{4}$, $\phi _{1}^{2}\phi _{2}$%
, $\phi _{2}^{2}$, $\phi _{1}\phi _{3}$, and $\phi _{4}$ all have $N$-ality $%
4$. Naively, the string tensions associated are $4\sigma _{1}$, $2\sigma
_{1}+\sigma _{2}$, $2\sigma _{2}$, $\sigma _{1}+\sigma _{3}$, and $\sigma
_{4}$, respectively. However, there is no symmetry principle prohibiting
mixing of operators of the same $N$-ality. If there is mixing, then only the
lightest string tension will be observed at large distances. Interactions
between modes will lead to such mixing, and the operators $Tr_{F}P^{k}$ will
exhibit a more complicated behavior.

Mixing between Polyakov loop operators of the same $N$-ality is easily
understood using character expansion techniques applied to lattice gauge
theories. For $d>2$, there are strong-coupling diagrams in which the sheet
between two Polyakov loops in a representation $R\,$split into a bubble with
sheets of representation $R_{1}$ and $R_{2}$. Such strong-coupling graphs
are non-zero when $R\subset R_{1}\otimes R_{2}$; a necessary but not
sufficient condition is that $R$ and $R_{1}\otimes R_{2}$ have the same $N$%
-ality. The same graph couples two representations $R$ and $R^{\prime }$ of
the same $N$-ality via the process $R\rightarrow R_{1}+R_{2}\rightarrow
R^{\prime }$. The case $d=2$ is special: in the continuum, there are no
gauge vector boson degrees of freedom, and the theory can be solved exactly,
yielding Casimir scaling. The lattice theory is also exactly solvable; it
reduces to a $d=1$ spin chain model, and there are no bubble diagrams as in
higher dimensions. Although the string tensions associated with a given
lattice action do not in general give Casimir scaling, 
the fixed-point lattice action in two dimensions
does yield results identical to the continuum\cite{Menotti:1981ry}.

\qquad It is easy to see that the $\left[ N/2\right] $ string tensions $%
\sigma _{1},\sigma _{2},..,\sigma _{\left[ N/2\right] }$ are all set
independently within the class of effective models. A \textbf{minimal} model
for the confined phase exhibiting this behavior is 
\begin{equation}
V=\sum_{k=1}^{\left[ N/2\right] }\frac{M_{k}^{4}}{k^{2}}%
Tr_{F}P^{k}Tr_{F}P^{+k} 
\end{equation}
$\,$where the $M_{k}$ are arbitrary. The $k$'th term in the sum forces $%
Tr_{F}P^{k}=0$, and gives rise to a mass for the mode $\phi _{k}$.

\vspace{1pt}Sine-law scaling arises naturally from a nearest-neighbor
interaction in the space of Polyakov loop eigenvalues. Consider the class of
potentials with pairwise interactions between the eigenvalues 
\begin{equation}
V_{2}=\sum_{j,k}v\left( \theta _{j}-\theta _{k}\right) . 
\end{equation}
as obtained, for example, 
by two-loop perturbation theory\cite{KorthalsAltes:1993ca}.
An elementary calculation shows that at tree level 
\begin{equation}
\sigma _{n}=\sqrt{\frac{2}{\kappa }\sum_{j=0}^{N-1}v^{\left( 2\right)
}\left( \frac{2\pi j}{N}\right) \sin ^{2}\left( \frac{\pi nj}{N}\right) } 
\label{eq:dispersion}
\end{equation}
where $v^{(2)}$ is the second derivative of $v$. This master formula relates
the string tensions to the underlying potential. It is essentially the
dispersion relation for a linear chain with arbitrary translation-invariant
quadratic couplings: nearest-neighbor, next-nearest neighbor, \textit{et
cetera}. If the sum is dominated by the $j=1$ and $j=N-1\,$terms,
representing a nearest-neighbor interaction in the space of eigenvalues,
then we recover sine-law scaling 
\begin{equation}
\sigma _{n}\simeq \sqrt{\frac{4}{\kappa }v^{\left( 2\right) }\left( \frac{%
2\pi }{N}\right) }\sin \left( \frac{\pi n}{N}\right) \rm{.} 
\end{equation}
There is a large class of potential which will give this behavior. $Z(N)$
scaling can be obtained by a very small admixture of other components of $%
v^{\left( 2\right) }$, as we show below.

The string tension associated with different $N$-alities has been measured
in $d=3$ and $4$ dimensions for $N=4$ and $6$, and in $d=4$ for $N=8$
\cite{Lucini:2001nv,Lucini:2002wg,DelDebbio:2001sj,DelDebbio:2003tk,Lucini:2004my}.
We examine the simulation data by inverting equation 
(\ref{eq:dispersion}) above to give a
measure of the relative strength of the couplings $v^{\left( 2\right)
}\left( 2\pi j/N\right) $. We normalize the result of this inversion such
that the sum of the independent couplings adds to one, and the results are
shown in Table \ref{Table1}. 
Sine-law scaling corresponds to a value of $1\,$for $j=1$, 
and $0$ for the other $\left[ N/2\right] -1$ independent couplings. For $%
Z(N)$ scaling, the large-$N$ limit gives $v^{\left( 2\right) }\left( 2\pi
j/N\right) \propto 1/j^{4}$, yielding the result shown in the table. Note
that the difference between sine-law scaling and $Z(N)$ scaling remains
small but finite, even as $N$ goes to infinity. The three-dimensional
simulation results clearly favor $Z(N)$ scaling. The two sets of
four-dimensional simulation results for $SU(4)$ and $SU(6)$ agree within
errors, but one set lies systematically closer to the sine-law predictions,
while the other set of results does not really favor either theoretical
prediction. The $SU(8)$ results nominally favor the sine-law prediction.
However, smaller error bars will be necessary to differentiate between
sine-law and $Z(N)$ scaling in four dimensions.

\begin{table}
\caption{{\label{Table1}}Relative strength of couplings $v^{\left( 2\right)}\left( 2\pi j/N\right) $}
\begin{tabular}{|l||l|l|l|l|l|}
\hline
& $d$ & $j=1$ & $j=2$ & $j=3$ & $j=4$ \\ \hline\hline
$SU(4)$\cite{Lucini:2001nv,Lucini:2002wg} & 3 & 0.957(6) & 0.043(2) &  &  \\ \hline
$SU(4)$\cite{Lucini:2004my} & 4 & 0.968(16) & 0.032(7) &  &  \\ \hline
$SU(4)$\cite{DelDebbio:2001sj,DelDebbio:2003tk} & 4 & 0.992(25) & 0.008(5) &  & \\ \hline
$SU(4)\,\,Z(N)$ & any & 0.9412 & 0.0588 &  &  \\ \hline
$SU(6)$\cite{Lucini:2001nv,Lucini:2002wg} & 3 & 0.930(5) & 0.065(8) & 0.004(5) &  \\ \hline
$SU(6)$\cite{Lucini:2004my} & 4 & 0.960(16) & 0.045(18) & -0.005(12) &  \\ \hline
$SU(6)$\cite{DelDebbio:2001sj,DelDebbio:2003tk} & 4 & 0.996(40) & 0.0003(218) & 0.004(26) &  \\ \hline
$SU(6)\,\,Z(N)$ & any & 0.9266 & 0.0618 & 0.0116 & \\ \hline
$SU(8)$\cite{Lucini:2004my} & 4 & 1.028(22) & -0.067(24) & 0.047(32) & -0.009(22) \\ \hline
$SU(8)\,\,Z(N)$ & any & 0.9249 & 0.0583 & 0.0130 & 0.0037 \\ \hline
$Z(N)$ $N\rightarrow \infty $ & any & 0.9239 & 0.0577 & 0.0114 & 0.0036 \\ \hline
sine Law & any & 1 & 0 & 0 & 0 \\ \hline
\end{tabular}
\end{table}

The ultimate origin of confinement influences the form of the potential
$V$, and the behavior of the different string tensions are derived in turn
from $V$.
In previous work on phenomenological models
of the gluon equation of state\cite{Meisinger:2001cq,Meisinger:2003id}, 
we considered models of the form 
\begin{equation}
f=-p=V(\theta )-2\int \frac{d^{3}k}{\left( 2\pi \right) ^{3}}Tr_{A}\ln
\left[ 1-Pe^{-\beta \omega _{k}}\right] 
\end{equation}
where $V$ is a phenomenologically chosen potential whose role is to favor
confinement at low temperature. We studied two physically motivated
potentials which reproduce $SU(3)$ thermodynamics well. Both give a
second-order deconfining transition for $SU(2)$, and a first-order
transition for $SU(N)$ with $N\geq 3$ in accord with simulation results. One
potential is a quadratic function of the eigenvalues 
\begin{equation}
V_{A}=v_{A}\sum_{\alpha =2}^{N}\sum_{\beta =1}^{\alpha -1}\left( \theta
_{\alpha }-\theta _{\beta }\right) \left( \theta _{\alpha }-\theta _{\beta
}-2\pi \right) 
\end{equation}
which appears as an $O(m^{2}T^{2})$ term in the high temperature expansions
at one-loop. This potential leads to $\sigma _{k}^{A}=\sigma _{1}$ for every 
$N$-ality. The other potential is the logarithm of Haar measure 
\begin{equation}
V_{B}=v_{B}\sum_{\alpha =2}^{N}\sum_{\beta =1}^{\alpha -1}\ln \left[ 1-\cos
\left( \theta _{\alpha }-\theta _{\beta }\right) \right] \rm{.} 
\end{equation}
\vspace{1pt}which is motivated by the appearance of Haar measure in the
functional integral. This term is cancelled out in perturbation theory for
flat space\cite{Weiss:1980rj}, but not in other geometries\cite{Aharony:2003sx}. The potential $V_{B}$ was first studied by Dyson in his
fundamental work on random matrices\cite{Dyson:1962}, and leads to 
\begin{equation}
\sigma _{k}^{B}=\sqrt{\frac{k\left( N-k\right) }{N-1}}\sigma _{1} 
\end{equation}
which one might call ''square root of Z(N) scaling''. Although these two
models are not consistent with the lattice simulation data for $\sigma _{k}$
for $N>3$, they remain viable phenomenological forms for $N=2$ and $3$. It
is interesting to note that there is a well-studied potential which gives $%
Z(N)$ scaling\cite{Calogero:1978}. It is the integrable 
Calogero-Sutherland-Moser potential 
\begin{equation}
V_{Z}\left( \theta \right) =\sum_{j\neq k}\frac{\lambda }{\sin ^{2}\left( 
\frac{\theta _{j}-\theta _{k}}{2}\right) } 
\end{equation}
which has been associated with
two-dimensional gauge theory\cite{Minahan:1993np,Polychronakos:1999sx}.

The effective action has domain wall solutions in the
deconfined phase, generalizing the behavior seen in the perturbative,
high-temperature form of the effective
action\cite{Bhattacharya:1990hk,Bhattacharya:1992qb}.
In this limit,
the $Z(N)$ symmetry breaks spontaneously, with $N$ equivalent vacua
characterized by $P=z^{n}I$.
There is a surface tension
associated with one-dimensional kink solutions of the effective equations
of motion which interpolate between  different phases.
There are $\left[ \frac{N}{2}\right]$ different surface tensions $\rho _{k}$,
each associated with the kink solution 
connecting the $n=0$ phase with the $n=k$ phase.
Giovannangeli and Korthals Altes\cite{Giovannangeli:2001bh}
have given an argument valid at
high temperature indicating that the surface tensions $\rho _{k}$ obey
$Z(N)$ scaling. This argument can be extended with minor modifications
to arbitrary potentials of the form $V_2$,
giving semiclassically
\begin{equation}
\rho _{k}=\frac{k\left( N-k\right) }{N-1}\rho _{1}. 
\end{equation}
in the entire deconfined phase.
%There are two inequivalent straight-line paths within the Cartan
%algebra. The first has the form 
%\begin{equation}
%P(x)=\exp \left[ \frac{2\pi i}{N}Y_{k}\,q(x)\right] 
%\end{equation}
%where $Y_{k}=diag\left[ k,..,k,N-k,..,N-k\right] $ with $N-k$ entries $k$
%and $k$ entries $N-k$\cite{Giovannangeli:2001bh}.
%The second is 
%\begin{equation}
%P(x)=\exp \left[ \frac{2\pi i}{N}kY_{1}\,q(x)\right] 
%\end{equation}
%For all potentials $V_{2}$ of the form given in eq. REF\ EQN, the first path
%leads to 
%\begin{equation}
%\rho _{k}=\frac{k\left( N-k\right) }{N-1}\rho _{1} 
%\end{equation}
%while the second leads to 
%\begin{equation}
%\rho _{k}=k\rho _{1} 
%\end{equation}
%and is not favored. Thus there is a semiclassical argument for $Z(N)$
%scaling of dual string tensions in the deconfined phase. 

A similar argument
has been given for the string tension in the confined phase of $2+1$%
-dimensional Polyakov models, which consist of gauge fields coupled to
scalars in the adjoint representation\cite{Kogan:2001vz}.
In this case, the effective potential has the form $V_{2}$ when written in
terms of dual variables. It is natural to speculate that there is a class of
self-dual effective models in two spatial dimensions with identical
string-tension scaling laws in both phases.

We have constructed the most general class of effective actions which can be
built from the eigenvalues of the Polyakov loop. $Z(N)$ symmetry requires a
symmetric distribution of eigenvalues in the confined phase, leading to a
special role for $Z(N)$ Fourier modes. Mode mixing provides a natural
crossover from Casimir scaling to scaling based on $N$-ality. There is a
natural association between sine-law scaling and strong nearest-neighbor
interactions between eigenvalues, and sine-law scaling and $Z(N)$ scaling
are in some sense quite close. Nevertheless, string tensions in different $N$%
-ality sectors are determined by the parameters of the effective action, and
are \textit{a priori} arbitrary. It is clear from studies of different
lattice gauge actions in two dimensions that the ratios of lattice string
tensions need not be universal. Only near the continuum limit are
two-dimensional string tension ratios universal. It remains mysterious why
the wide class of models so far studied have not provided us with a wider
variety of string tension scaling laws. We do not know if additional fields
of zero N-ality can change string tension scaling in the continuum, and
there has
not been much exploration of the possible effect of fields that preserve a
non-trivial subgroup of $Z(N)$. Studies of such models might give insight
into the connection between confinement mechanisms and string-tension
scaling laws.

\begin{acknowledgments}
MCO is grateful to the U.S. Dept. of Energy for financial support.
\end{acknowledgments}

\end{document}